\documentclass[12pt,twoside]{article}
\usepackage{fleqn,espcrc1,epsfig}

\usepackage{graphicx}
\usepackage[figuresright]{rotating}

\def\Journal#1#2#3#4{{#1} {\bf #2}, #3 (#4)}
\def\AoP{\em Ann. of Phys.}

\def\NIMA{{\em Nucl. Instrum. Methods} A}
\def\NPA{{\em Nucl. Phys.} A}
\def\NPB{{\em Nucl. Phys.} B}
\def\PLB{{\em Phys. Lett.}  B}
\def\PRL{\em Phys. Rev. Lett.}
\def\PRC{{\em Phys. Rev.} C}

\def\ZPA{{\em Z. Phys.} A}
\def\EPJA{{\em Eur. Phys. J.} A}

\def\JPG{\em J. Phys. G: Nucl. Part. Phys.}

\newcommand{\Den}{$D(\vec e,e'\vec n)$~}
\newcommand{\Hen}{$^{3}$$\vec{He}(\vec e,e'n)$~}

\newcommand{\AmS}{{\protect\the\textfont2
  A\kern-.1667em\lower.5ex\hbox{M}\kern-.125emS}}

\title{Double polarization experiments at intermediate energy}

\author{H. Schmieden\address{Institut f\"ur Kernphysik,
                             Johannes Gutenberg-Universit\"at \\ 
                             J.J. Becher-Weg 45,
                             55099 Mainz, Germany}%
        }
       
\begin{document}

\maketitle

\begin{abstract}
At modern electron accelerators with highly polarized, intense, high duty 
factor beams double polarization coincidence experiments became 
feasible with good statistical accuracy. 
The strong potential towards the precise determination of small
nucleon structure quantities  is illustrated by two recent examples from MAMI.
The measurement of $G_E^n$  in the quasifree reaction $D(\vec e, e'\vec n)p$
lead to a new parametrization of $G_E^n$  which is significantly above the
previously preferred one from elastic $e-D$  scattering.
A $p(\vec e, e'\vec p)\pi^0$  experiment at the energy of the  
$\Delta$ resonance yields preliminary results for the longitudinal
quadrupole mixing.
Both experimental errors and model uncertainties are complementary to 
unpolarized measurements.
\end{abstract}

\section{INTRODUCTION}

Polarization experiments offer the possibility 
to measure interferences of different amplitudes. 
This is particularly interesting in the situation of 
a small quantity in the vicinity of a dominating large one.
Due to the insensitivity of polarization observables to many calibration
factors the small quantities can be reliably extracted.

For these reasons a variety of experimental programs in the intermediate 
energy range has been established at the electron accelerator facilities
ELSA, MAMI, MIT-Bates, NIKHEF and TJNAF. 
In the following, two recent examples of double polarization experiments at 
MAMI will be discussed:
the measurement of the neutron electric form factor, $G_E^n$, in the 
quasifree $D(\vec e, e'\vec n)p$  reaction, and the extraction of the
longitudinal quadrupole mixing in the N to $\Delta$  transition from 
$p(\vec e, e'\vec p)\pi^0$.  

Both experiments require longitudinally polarized, high duty factor beams
in combination with recoil polarimetry. 
The nucleons are detected in parallel kinematics, i.e. along the direction
of the momentum transfer, $\vec q$.
The cartesian components of the nucleon polarization, $P_{x,y,z}$,  
are advantageously expressed in the plane of incoming ($\vec k_i$) 
and scattered electron ($\vec k_f$):
$  \hat{y} = (\vec k_i \times \vec k_f) / |\vec k_i \times \vec k_f|  $,
$  \hat{z} = \vec q / |\vec q|  $, and
$  \hat{x} = \hat{y} \times \hat{z}  $.

\section{MEASUREMENT OF \boldmath${G_E^n}$  via 
                        \boldmath$D(\vec e, e'\vec n)p$}

Double polarization observables in quasifree electron-deuteron scattering
offer high sensitivity to 
$G_E^n$ due to an interference with the large magnetic form factor, $G_M^n$, 
combined with negligible dependence on the deuteron wavefunction
\cite{Arenhoevel}.
For the ideal case of free electron-neutron scattering, $n(\vec e, e' \vec n)$,
Arnold, Carlson and Gross \cite{ACG81} obtained for the components of the
recoil polarization\footnote{Equivalently, the scattering of longitudinally 
polarized electrons off a polarized neutron target leads to a cross section 
asymmetry with regard to reversal of beam helicity.} 
\begin{eqnarray}
P_x &=& - P_e \frac{\sqrt{2\tau\epsilon(1-\epsilon)} G_E^n G_M^n}
                   {\epsilon (G_E^n)^2 + \tau (G_M^n)^2}            \\
P_y &=& 0                                                       \\
P_z &=& P_e \frac {\tau\sqrt{1-\epsilon^2} (G_M^n)^2}
                  {\epsilon (G_E^n)^2 + \tau (G_M^n)^2}.                
\end{eqnarray}
$\epsilon = ( 1 + \frac{2|\vec q|^2}{Q^2} \tan^2{\frac{\vartheta_e}{2}} )^{-1}$
is the photon polarization parameter,
$Q^2 = -q_{\mu}q^{\mu}$  the squared four momentum transfer
and $\tau = Q^2/4m_n^2$  represents the momentum transfer in units of 
the neutron mass, $m_n$.
$\vartheta_e$  denotes the electron scattering angle,
and $P_e$  the longitudinal polarization of the electron beam.
\begin{figure}[t]
\begin{center}
   \epsfig{file=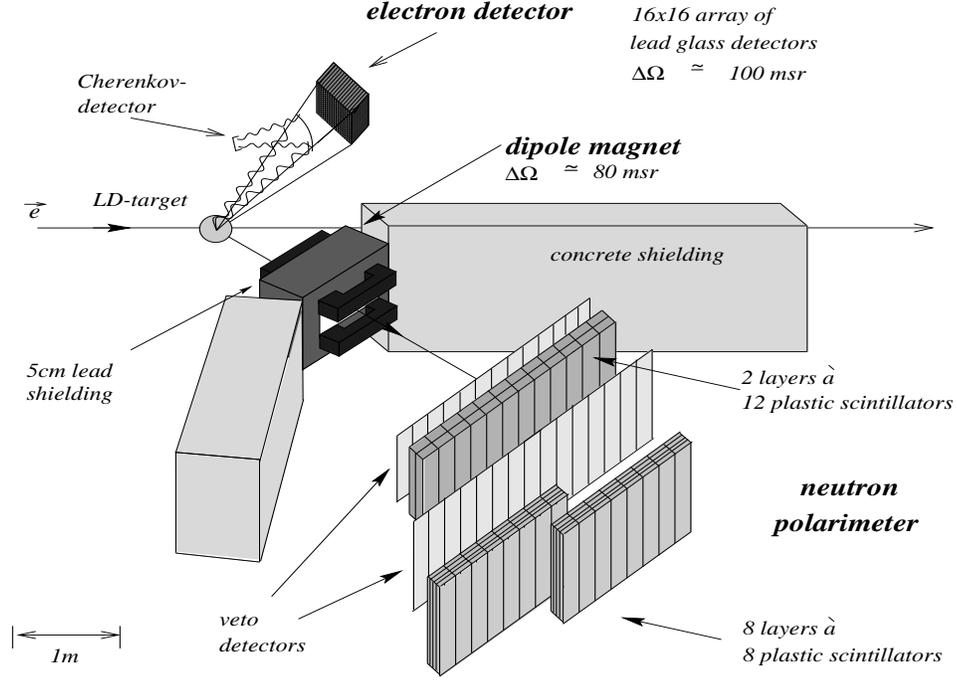, width=12.5cm, height=9cm}
   \caption{Setup of the $D(\vec e, e'\vec n)$  experiment at MAMI}
   \label{fig:gen_setup}
\end{center}
\end{figure}

The longitudinally polarized electron beam 
($I \simeq 2.5\,\mu$A, $P_e \simeq 75$\,\%)
hit a 5\,cm long liquid deuterium target
and the scattered electrons were detected in a 256 element lead glass 
array (Fig.\ref{fig:gen_setup}). 
The energy resolution of $\delta E / E \simeq 25$\,\% was sufficient to
suppress pion production events. 
Only the electron angles, which were measured with an accuracy of 
$\delta \vartheta, \delta \phi \simeq 3.5$\,mrad entered the 
event reconstruction,
which became kinematically complete through the measurement of the
neutrons time-of-flight and hit position in the front plane of
the neutron detector.
\begin{figure}[htb]
\begin{minipage}[t]{80mm}
\epsfig{file=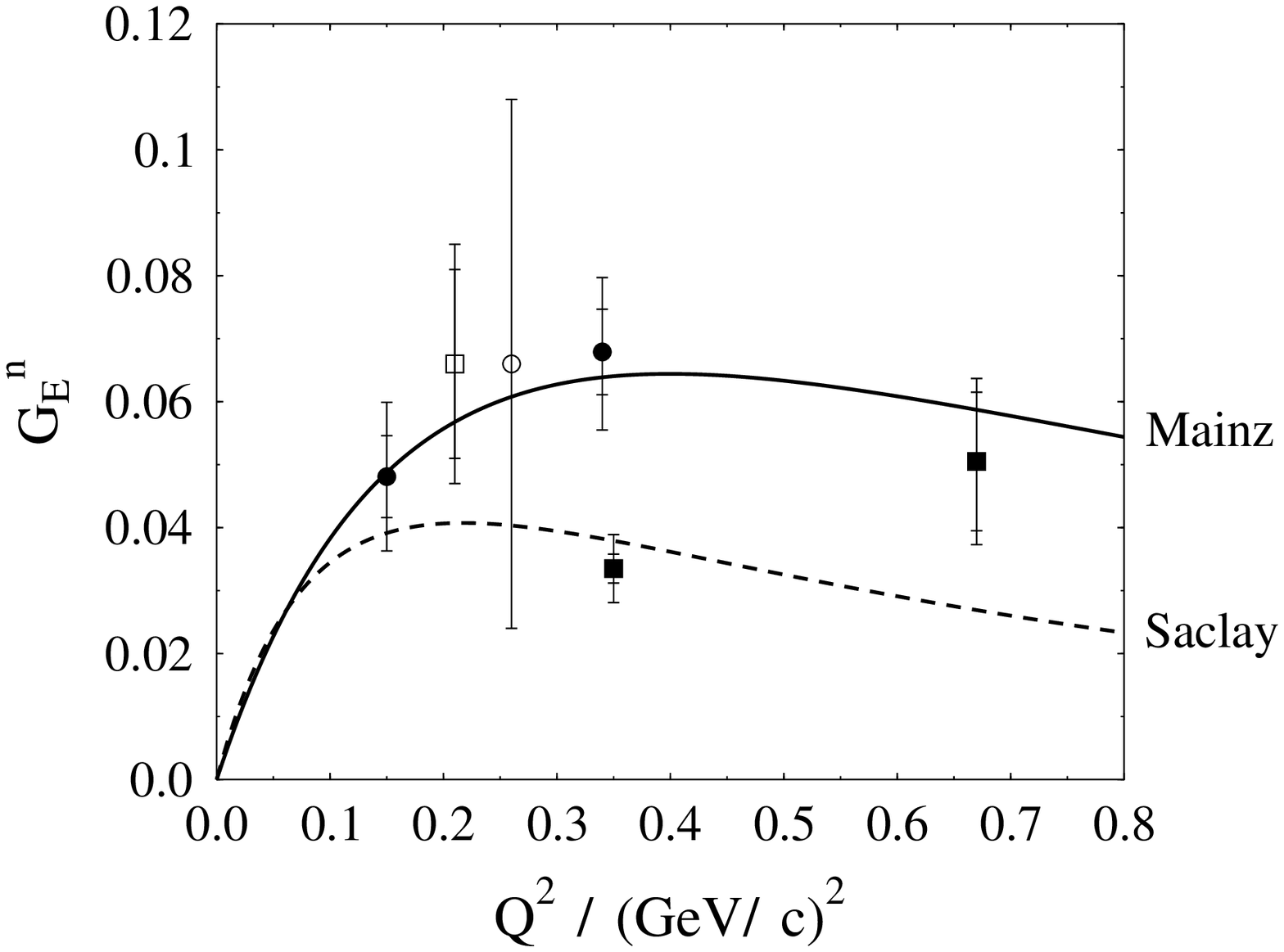, width=8.5cm}
\caption{Results for $G_E^n$  from double polarization experiments.
          The data points and curves are discussed in the text. }
\label{fig:gen}
\end{minipage}
\hspace{\fill}
\begin{minipage}[t]{75mm}
    \epsfig{file=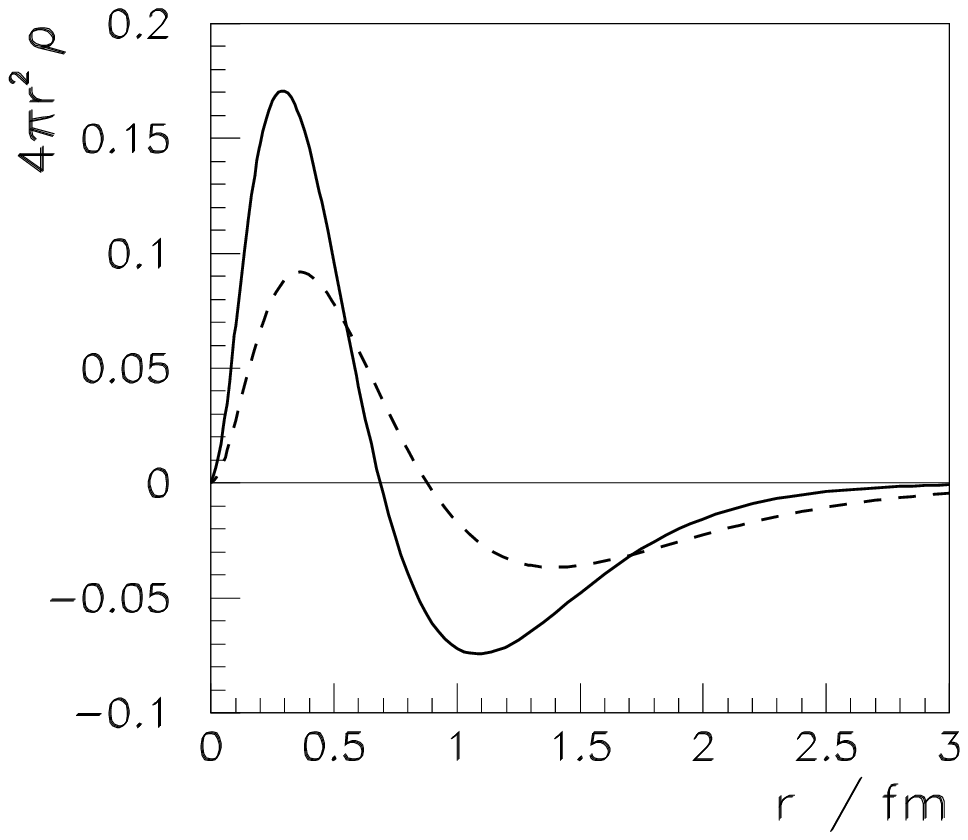, width=7.5cm}
\caption{Neutron charge distribution for  
         the Mainz (full) and the preferred
         Saclay (broken) $G_E^n$  parametrization (see text). }
\label{fig:charge}
\end{minipage}
\end{figure}

The neutron polarization can be analyzed in the detection process 
itself \cite{Taddeucci85}.
This required a second neutron detection in one of 
the rear detector planes,
which yielded the polar and azimuthal angles, $\Theta_n'$  and $\Phi_n'$,
of the analyzing scattering in the front wall.
With the number of events $N^{\pm}(\Phi_n')$  for $\pm$  helicity states of
the electron beam the azimuthal asymmetry, $A(\Phi_n')$, 
was determined through the ratio
\begin{equation}
\frac{1-A(\Phi_n')}{1+A(\Phi_n')} =
   \sqrt{ \frac{ N^+(\Phi_n') \cdot N^-(\Phi_n'+\pi) }
               { N^-(\Phi_n') \cdot N^+(\Phi_n'+\pi)} } ,
\end {equation}
which is insensitive to detector efficiencies and luminosity variations.
The extraction of $P_x$  from 
$A(\Phi_n') = \epsilon_{\mbox{\scriptsize eff}} \cdot P_x \cdot \sin\Phi_n'$  
requires the calibration of the effective analyzing power, 
$\epsilon_{\mbox{\scriptsize eff}}$, of the polarimeter.
This, however, varies strongly with the event composition as determined
by hardware conditions during data taking and software cuts applied in
the offline analysis.

For the first time, the problem of calibration of the effective analyzing 
power has been avoided by controlled precession of the neutron spins in the
field of a dipole magnet in front of the polarimeter
\cite{Ostrick99}. 
After precession by the angle $\chi$  the transverse neutron polarization
behind the magnet, $P_{\perp}$, is a superposition of $x$  and $z$
components, and likewise is the measured asymmetry:
\begin{equation}
A_{\perp} = A_x \cos\chi - A_z\sin\chi.
\end{equation}
For the particular case of the zero crossing, $A_{\perp}(\chi_0) = 0$,
one immediately gets the relation
\begin{equation}
\tan\chi_0 = \frac{A_x}{A_z} = \frac
{\epsilon_{\mbox{\scriptsize eff}} \cdot 
 P_e\cdot\sqrt{2\tau\epsilon(1-\epsilon)}\,G_E^n\cdot G_M^n}
{\epsilon_{\mbox{\scriptsize eff}} \cdot 
 P_e\cdot\tau\sqrt{1-\epsilon^2}\,(G_M^n)^2 }.
\end{equation}
Obviously, this ratio is independent of both $P_e$  and
$\epsilon_{\mbox{\scriptsize eff}}$  and
it directly yields $G_E^n/G_M^n$.
The neutron electric form factor has been extracted 
relying on the dipole values for $G_M^n$.

Data have been taken around $Q^2 = 0.32$\,(GeV/c)$^2$  and 
$Q^2 = 0.12$\,(GeV/c)$^2$.
According to calculations of H. Arenh\"ovel 
the effect of final state interaction on $P_x$ 
-- which is dominated by charge exchange of the outgoing nucleons --
is almost negligible at the higher momentum transfer.
However, due to the small relative energy in the n-p final state,
it becomes important at the small $Q^2$.
For the MAMI \Den data the influence of FSI has been explicitly studied for 
the first time \cite{Herberg99}.
A 100\,\% correction is required at $Q^2 = 0.12$\,(GeV/c)$^2$  which drops to
8\,\% at $Q^2 = 0.35$\,(GeV/c)$^2$.
The corrected results are depicted in Fig.\ref{fig:gen} 
as full circles with 
statistical (inner) and systematical (outer) error.
A FSI correction is implicitly also included in the NIKHEF result
\cite{Passchier99} (open square), 
but not in the Bates one \cite{Eden94} (open circle).
The full squares represent the uncorrected MAMI results for \Hen. 
At $Q^2 = 0.6$\,(GeV/c)$^2$  \cite{Rohe99} FSI is expected to be negligible 
due to the large kinetic energy of the ejected neutron.
However, first, still incomplete 3-body calculations 
indicate a substantial correction of the data point at 
$Q^2 = 0.36$\,(GeV/c)$^2$  (full square)\cite{Becker99} 
towards larger $G_E^n$.
Excluding this data point, a new fit of the dipole-ansatz 
\begin{equation}
G_E^n = - \frac{\mu_n \tau}{1 + \eta \tau} \cdot 
          \left[ 1 + \frac{Q^2}{0.71\,(GeV/c)^2} \right]^{-2}
\end{equation}
to the recent MAMI double polarization results yielded $\eta = 3.4$.
$\mu_n$  is the neutron magnetic moment.
As shown in Fig.\ref{fig:gen}, this fit
lies almost a factor of two above the previously favoured result from
elastic $D(e,e')$  scattering, where the Paris potential has been used for the
unfolding of the wave function contribution \cite{Platchkov90}.
This causes a significant difference of the neutron charge distributions 
as obtained by Fourier transformation of $G_E^n$.
The full and broken curve in Fig.\ref{fig:charge} correspond to the
Mainz and Saclay $G_E^n$  parametrizations, respectively.

\section{RECOIL POLARIZATION IN THE \boldmath$p(\vec e, e'\vec p)\pi^0$  
         REACTION AND THE C2/M1 RATIO IN THE N TO 
         \boldmath$\Delta$  TRANSITION}

The spherical symmetry of the distributions of charge and magnetism 
within the nucleon was discussed since the early days
of the quark model \cite{BM65}, although
the nucleon's spin 1/2 forbids the existence of a static
quadrupole moment.
However, in the 
$N \rightarrow \Delta_{33}(1232)$  transition
quadrupole components are allowed aside of the dominant
spin-flip of the unaligned constituent quark. 
In the almost exclusive decay of the $\Delta_{33}(1232)$  resonance into the 
$N \pi$  channel, this small quadrupole mixing is associated with 
small, but non-zero, 
electric quadrupole to magnetic dipole (EMR) and Coulomb quadrupole to
magnetic dipole ratios (CMR).
These are defined as:
\begin{eqnarray}
\mbox{EMR} &=& \Im m \{E_{1+}^{3/2}\} / \Im m \{M_{1+}^{3/2}\}  
\label{eq:EMR} \\
\mbox{CMR} &=& \Im m \{S_{1+}^{3/2}\} / \Im m \{M_{1+}^{3/2}\},
\label{eq:CMR}
\end{eqnarray}
where the pion multipoles, $A^I_{l_{\pi}\pm}$, are characterized through their 
magnetic, electric or longitudinal (scalar) nature, $A$, 
the isospin, $I$, and the pion-nucleon relative angular momentum, $l_{\pi}$, 
whose coupling with the nucleon spin 
is indicated by $\pm$.

While the EMR has recently been measured at the photon point ($Q^2 = 0$) 
\cite{Beck97,Blanpied97},
the determination of the 
longitudinal quadrupole mixing requires
pion electroproduction experiments.
Double polarization observables in the $p (\vec e, e' \vec p) \pi^0$  reaction
offer high sensitivity to the CMR.   
In parallel kinematics, the components of the proton polarization
are given in s- and p-wave approximation by \cite{HS98} 
\footnote{In contrast to ref.\cite{HS98} here the conventions of 
          ref.\cite{DT92} are used.
          The different p-wave signs compared to \cite{HS98} are due to
          an inconsistency in \cite{RD89}.}:
\begin{eqnarray}
\sigma_0 P_x &=& P_e \cdot 
                 c_- \cdot
                 \Re{e}\{(4 S_{1+} + S_{1-} - S_{0+})^* \cdot
                         (M_{1+} - M_{1-} - E_{0+} + 3E_{1+})\}
\label{eq:P_x_sp} \\
\sigma_0 P_y &=& 
                 c_+ \cdot
                 \Im{m}\{(4 S_{1+} + S_{1-} - S_{0+})^* \cdot
                         (M_{1+} - M_{1-} - E_{0+} + 3E_{1+})\}
\label{eq:P_y_sp} \\
\sigma_0 P_z &=& P_e \cdot \sqrt{1-\epsilon^2} \cdot
                 (
                 |M_{1+}|^2 + |M_{1-}|^2 + 9|E_{1+}|^2 + |E_{0+}|^2 + 
                 \nonumber \\ & &
                 + \Re{e}\{6E_{1+}^*(M_{1+}-M_{1-}) - 2M_{1+}^*M_{1-}
                       - 2E_{0+}^*(M_{1+}-M_{1-}+3E_{1+}) \}
                 ),
\label{eq:P_z_sp} 
\end{eqnarray}
where 
$c_{\pm}=\sqrt{2\epsilon_L(1\pm\epsilon)}\frac{\omega_{cm}}{|\vec q_{cm}|}$.
$\omega_{cm}$  and $\vec q_{cm}$  denote the energy and momentum transfer 
in the cm-frame and $\epsilon_L = \frac{Q^2}{\omega_{cm}^2}\epsilon$.
The $p-\pi^0$  multipoles of Eqs.\ref{eq:P_x_sp}-\ref{eq:P_z_sp}, 
$ A_{l\pm}$, are related to the isospin multipoles by
$ A_{l\pm} = A_{l\pm}^{1/2} + \frac{2}{3} A_{l\pm}^{3/2} $.

As a consequence of the $M_{1+}^{3/2}$  dominance, $\Re{e}M_{1+}$ vanishes
very closely to the resonance position ($W=1232$\,MeV), i.e. 
\begin{eqnarray}
\sigma_0 P_x &=& P_e \cdot 
                 c_- \cdot
                 4 \Im{m}S_{1+} \Im{m}M_{1+} + n.l.o.  
\label{eq:P_x_simple}                                      \\
\sigma_0 P_y &=& 
                 c_+ \cdot
                 4 \Re{e}S_{1+} \Im{m}M_{1+} + n.l.o.  
\label{eq:P_y_simple}                                      \\
\sigma_0 P_z &=& P_e \cdot \sqrt{1-\epsilon^2} |M_{1+}|^2 + n.l.o.
\label{eq:P_z_simple}
\end{eqnarray}
$S_{1+}$, $E_{1+}$  and $M_{1+}$  are the only multipoles which couple 
to the $\Delta$  resonance, i.e. possess significant imaginary parts.
With a purly Born - i.e. real - background the imaginary parts of the other
multipoles vanish.
Rescattering effects produce only small imaginary parts in the $S_{0+}$  
and the $E_{0+}$  amplitudes.
Therefore the higher order contributions, $n.l.o.$,  to $P_x$ 
(Eq.\ref{eq:P_x_simple}) can expected to be small.
The situation is different for $P_y$, because here all the real Born multipoles
contribute (Eq.\ref{eq:P_y_sp}).
From Eqs.\ref{eq:P_x_simple} and \ref{eq:P_z_simple} it is evident that
the CMR can be almost directly determined from the 
polarization ratio 
$ R = \frac{\sqrt{1-\epsilon^2}}{4 c_-} \frac{P_x}{P_z} $ \cite{HS98}. 

At the 3-spectrometer setup \cite{Blomqvist98} of the Mainz microtron MAMI a
$p(\vec e, e'\vec p) \pi^0$  experiment with longitudinally polarized electron 
beam and measurement of the recoil proton polarization has been performed.
At $Q^2 = 0.121$\,(GeV/c)$^2$  a range of invariant energies of 
$W = 1200 ... 1260$\, MeV was covered.
\begin{figure}[t]
\begin{center}
   \epsfig{file=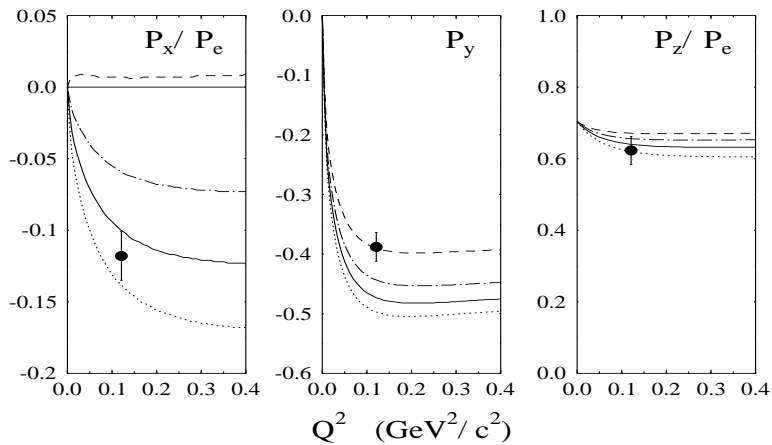, width=11cm, height=7cm}
   \caption{ Preliminary results for the measured polarization components.
             The dashed, dot-dashed, full and dotted
             curves correspond to calculations
             within the UIM for  
             CMR$=0$, $-2.9$, $-4.8$, $-6.7$\,\%, respectively. }
   \label{fig:polarizations}
\end{center}
\end{figure}

The proton polarization was measured through the standard technique of 
inclusive p - $^{12}$C scattering.
Therefore the detector package of spectrometer A was supplemented by
a 7\,cm thick carbon scatterer followed by two double planes of horizontal
drift chambers \cite{Pospischil99}. 
This setup allowed the eventwise reconstruction of the proton carbon 
scattering angles $\Theta_C$  and $\Phi_C$  with a resolution of 3\,mrad.
The azimuthal modulation of the cross section
\begin{equation}
\sigma_C = \sigma_{C,0} [ 
           1+A_C(P_y^{fp}\cos\Phi_C-P_x^{fp}\sin\Phi_C) ]
\end{equation} 
is related to the proton polarization;
$\sigma_{C,0}$  denotes the polarization independent part of the inclusive 
cross section and $A_C$  the analyzing power, which was parameterized according
to \cite{McNaughton85}.

The two polarization components $P_x^{fp}$  and $P_y^{fp}$  are
measured behind the spectrometer's focal plane. 
It is possible to determine
all three components at the electron scattering vertex due to the spin 
precession in the spectrometer and the redundancy provided by the helicity 
flip of the electron beam.

The spin precession along the proton trajectories through the magnetic 
fields of the QSDD spectrometer A \cite{Blomqvist98} was computed by
stepwise numerical integration of the BMT-equation. 
On the basis of several thousand rays which were distributed over the
large spectrometer acceptance a 5-dimensional 
spin precession matrix describing the spectrometer's 
`polarization-optics' was generated \cite{Pospischil99}.
The results were checked through the elastic scattering reaction 
$p(\vec e, e'\vec p)$     
where the proton polarization is determined by electron kinematics and the 
proton elastic form factors.  

The matrix was then used for the extraction of the recoil proton 
polarization in the $p(\vec e, e'\vec p)\pi^0$  reaction.
The preliminary results for the three polarization components are shown in 
Fig.\ref{fig:polarizations}. 
The curves represent full (i.e. without restriction to s- and p-waves) 
calculations in the framework of the Mainz unitary isobar model 
(UIM) \cite{Drechsel99}. 
$P_x$  is very sensitive to the CMR and from this polarization component
a value of 
$\mbox{CMR} = (-5.19 \pm 0.75_{stat} \pm 0.60_{syst})$\,\% is extracted.
However, since the influence of systematic uncertainties is reduced,
it is preferrable to extract the CMR from the ratio 
$R = (-6.17 \pm 0.99 \pm 0.56)$\,\%.
The preliminary result for the full UIM analysis of the ratio yields 
$\mbox{CMR} = (-5.25 \pm 0.82 \pm 0.48)$\,\%.
\begin{figure}[t]
\begin{center}
   \epsfig{file=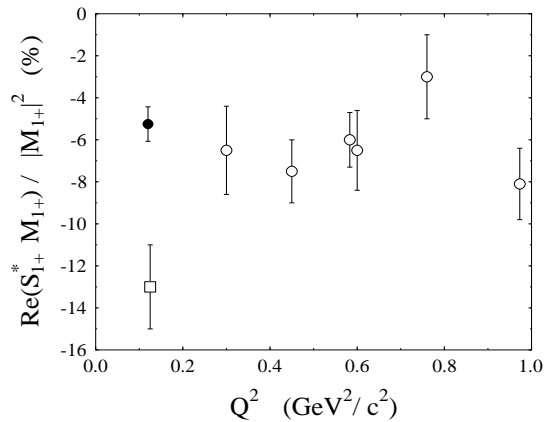, width=8.5cm}
   \caption{Preliminary
            result of the CMR as extracted from $P_x/P_z$  of this experiment
            (full circle) in comparison with unpolarized measurements 
            from
            DESY, NINA, the Bonn synchrotron \protect{\cite{DESY-NINA}} 
            (open circles), 
            and ELSA \protect{\cite{Kalleicher97}} (open square).
            Errors are purely statistical.  }
   \label{fig:s1_comparison}
\end{center}
\end{figure}

Although the CMR is determined in the $\pi^0$  channel,
at the resonance position 
(defined by $\Re{e}M^{3/2}_{1+}=0$, i.e. at $W = 1232$\,MeV)
it is believed to come very close to the value for the isospin-3/2 channel 
(Eq.\ref{eq:CMR}) due to resonance dominance.  
In the UIM the $P_x$-interference in the $\pi^0$  channel 
practically equals at resonance the isospin-3/2 one:
$ \Re{e}\{S^{\pi^0 *}_{1+} M^{\pi^0}_{1+}\}/|M^{\pi^0}_{1+}|^2 =
  \Re{e}\{S^{3/2 *}_{1+} M^{3/2}_{1+}\}/|M^{3/2}_{1+}|^2 $.

In Fig.\ref{fig:s1_comparison} the preliminary result is plotted along 
with published results from 
coincident $\pi^0$  experiments with unpolarized electrons.
It is compatible with these data \cite{DESY-NINA}
and with recent, yet unpublished results \cite{Gothe,Mertz99},
which are not included in Fig.\ref{fig:s1_comparison}. 
However, one Bonn result \cite{Kalleicher97} seems to be incompatible
with all other data. 
Wether this is an experimental or statistical artefact, or a severe hint
that there is another than expected $\Im{m}S_{0+}$  background contribution
- which enters with different sign in the kinematics of that experiment 
compared to all other ones -
is still an open problem but hoped to be decided soon
\cite{HS98a}.

\section{SUMMARY}

Two examples of double polarization experiments from MAMI in the 
medium energy range have been presented: 
quasifree $D(\vec e, e'\vec n)p$  scattering and 
$p(\vec e, e'\vec p)\pi^0$  in the energy range of the $\Delta$  
resonance.
These experiments resemble each other in that the recoiling nucleons
are detected in coincidence with the scattered electrons in the direction 
of the momentum transfer and the nucleon polarization is analyzed in
appropriate polarimeters.
In both cases the interesting observables can be extracted from ratios
of polarizations with the advantage of a particular insensitivity to 
calibration factors.

The new double polarization experiments in quasifree kinematics establish
that the neutron electric form factor, $G_E^n$, is almost a
factor of two higher than the previously favoured result of elastic
$e-D$  scattering. 
This corresponds to a significant different charge distribution
inside the neutron.

The recoil polarization observables in the pion electroproduction reaction
are sensitive to the longitudinal quadrupole mixing, CMR,  in the 
$p \rightarrow \Delta^+$  transition.
The preliminary result seems to support older unpolarized measurements 
within their larger errors. 
The discrepancy with one recent experiment is possibly due to a remaining 
model dependence in the analyses of this type of measurements.
In this case the double polarization observables contribute information
which is complementary to unpolarized measurements concerning the separation
of the non-resonant background contributions.

\section{ACKNOWLEDGEMENTS}

The experiments have been performed within the collaborations
A1 and A3 at MAMI with contributions from institutes of the universities of
Basel, Bonn, Glasgow, Ljubljana, Mainz, Rutgers and T{\"u}bingen.
The preliminary CMR-results are based on the doctoral thesis of Th. Pospischil.
Calculations have been provided by H. Arenh{\"o}vel concerning the 
influence of FSI on the polarization observables in quasielastic $e-D$  
scattering, and by S. Kamalov and L. Tiator for the extraction of the 
CMR from the $\pi^0$  experiment.
Financial support came from the Deutsche Forschungsgemeinschaft
(SFB\,201).

\end{document}